\documentclass[%
 reprint,
 amsmath,amssymb,
 aps,
]{revtex4-2}

\usepackage{graphicx}
\usepackage{dcolumn}
\usepackage{bm}
\usepackage{hyperref}

\usepackage{amsthm}
\usepackage{amsmath}
\usepackage{ amssymb }
\usepackage{ mathrsfs }
\usepackage{physics}
\usepackage{tikz-cd}

\newcommand{\nc}{\newcommand}
\nc{\lb}{\llbracket}
\nc{\rb}{\rrbracket}
\nc{\gl}{\llbracket}
\nc{\gr}{\rrbracket}
\nc{\bbR}{\mathbb{R}}
\nc{\bbC}{\mathbb{C}}
\nc{\bbZ}{\mathbb{Z}}
\nc{\cO}{\mathcal{O}}
\nc{\cS}{\mathcal{S}}
\nc{\cM}{\mathcal{M}}
\nc{\cT}{\mathcal{T}}
\nc{\cX}{\mathcal{X}}
\nc{\cQ}{\mathcal{Q}}
\nc{\cD}{\mathcal{D}}
\nc{\cC}{\mathcal{C}}
\nc{\cL}{\mathcal{L}}
\nc{\cG}{\mathcal{G}}
\nc{\cF}{\mathcal{F}}
\nc{\cI}{\mathcal{I}}
\nc{\cN}{\mathcal{N}}
\nc{\pd}{\partial}
\nc{\la}{\lambda}

\DeclareMathOperator{\arccosh}{arcCosh}

\newtheorem{thr}{Theorem}[section]
\newtheorem{prop}{Proposition}[section]

\makeatother

\begin{document}

\preprint{APS}

\title{Tame Embeddings, Volume Growth, and Complexity of Moduli Spaces}

\author{Thomas W.~Grimm}
\author{David Prieto}
\author{Mick van Vliet}%

\affiliation{%
Institute for Theoretical Physics, Utrecht University,
\\
Princetonplein 5, 3584 CC Utrecht, 
The Netherlands 
}%


\begin{abstract}
\noindent

Quantum gravity is expected to impose constraints on the moduli spaces of massless fields that can arise in effective quantum field theories. A recent proposal asserts that the asymptotic volume growth of these spaces is severely restricted, and related to the existence of duality symmetries. In this work we link this proposal to a tameness criterion, by suggesting that any consistent moduli space should admit a tame isometric embedding into Euclidean space. This allows us to promote the volume growth constraint to a local condition, and give the growth coefficient a geometric interpretation in terms of complexity. We study the implications of this proposal for  the emergence of dualities, as well as for the curvature and infinite distance limits of moduli spaces. 
\end{abstract}

\maketitle

\section{\label{sec:intro}Introduction}

Finiteness is one of the key principles that ties together many of the Swampland conjectures used to characterize the effective field theories compatible with quantum gravity \cite{Vafa:2005ui}. For instance, it is present in the No Global Symmetries Conjecture and the Weak Gravity Conjecture, constraining the number of black hole remnants and, more broadly, is setting bounds on the entropy of any quantum gravity system \cite{Arkani-Hamed:2006emk,Banks:2010zn}. It is also essential in the Distance Conjecture, prohibiting the effective field theory from accessing regions of infinite distance in moduli space \cite{Ooguri:2006in} and thus keeping the number of light degrees of freedom finite. It even seems to be a defining feature of the String Landscape itself, since all evidence has led to conjecture that the number of effective theories valid below a fixed cut-off scale that are consistent with quantum gravity is finite \cite{Douglas:2003um, Vafa:2005ui, 
Acharya:2006zw, Hamada:2021yxy}. 

In the context of finiteness, great progress has been accomplished by the mathematical community during the last three decades, leading to the creation of a new mathematical notion, known as tameness or o-minimality, which can be used to answer finiteness questions in various fields, ranging from number theory to geometry. Its potential to improve our understanding of the Swampland was first discussed in \cite{Grimm:2021vpn} and later in \cite{Douglas:2022ynw,Douglas:2023fcg}. In these works, it was conjectured that all effective theories compatible with quantum gravity are defined in terms of tame spaces and tame coupling functions. Tameness has been used to prove the finiteness of self-dual vacua satisfying the tadpole bound in F-theory compactifications \cite{Bakker:2021uqw}, see also \cite{Grimm:2021vpn,Grimm:2023lrf}, to sharpen our understanding of the distance conjecture \cite{Grimm:2022sbl}, and to suggest a notion of complexity for quantum field theories \cite{Grimm:2023xqy,Grimm:2024elq}.

 It is thus natural to apply this framework to other questions related with the Swampland. In this short note we specifically aim to use tameness arguments to derive bounds for the volume growth of moduli spaces in quantum gravity, recovering the compactifiability condition presented in \cite{Delgado:2024skw}. There, it is argued that demanding the finiteness of the number of massless states obtained upon compactifying all spatial dimensions implies that the associated moduli space must be compactifiable. In this context, a moduli space $\mathcal{M}$ with geodesic distance function $d(\cdot\, ,\, \cdot)$ is said to be compactifiable if for any $\phi_0\in \mathcal{M}$ and $\cD>0$ the set 
\begin{equation}
    \mathcal{M}_\mathcal{D} (\phi_0)=\{\phi\in \mathcal{M} \ | \ d(\phi,\phi_0)\leq \mathcal{D}\}\,,
\end{equation}
satisfies the volume growth condition
\begin{equation}
    {\rm Vol}(\mathcal{M}_\mathcal{D})\ll \mathcal{D}^{n+\epsilon}\,,
    \label{eq: compacty condition}
\end{equation}
for arbitrary $\epsilon>0$ and $n={\rm dim}(\mathcal{M})$ in the asymptotic limit $\mathcal{D}\rightarrow\infty$. In other words, the volume of a geodesic ball in the moduli space should grow no faster than the volume of a Euclidean ball.
 
 Note that this conjecture was formulated as an asymptotic statement. It puts non-trivial constraints on effective theories, since according to the original formulation of the Distance Conjecture in \cite{Ooguri:2006in}, every moduli space has points at infinite distance boundaries and the $\cD \rightarrow \infty$ limit exists. However, the compactifiability condition is  trivially satisfied for moduli spaces with finite volume. This is, for example, the case for complex structure moduli spaces Calabi-Yau manifolds \cite{Lu:2005bj} and their associated effective theories. It continues to hold in known infinite volume examples, such as the moduli spaces  of M-theory on a Klein bottle and type IIA supergravity, whose volume grows like the Euclidean space as a function of the distance \cite{Delgado:2024skw}.

The main focus of this note is to derive a sharp local bound on the volume growth as a function of the geodesic distance for moduli spaces that admit a tame isometric embedding into Euclidean space. In particular, this bound implies the compactifiability of the associated space and leads us to conjecture that tame isometric embeddability is a general feature of quantum gravity moduli spaces, refining  the tameness conjecture proposed in \cite{Grimm:2021vpn}. As we will see, the important role of the embedding is a manifestation of the properties of the relevant functions defined over the moduli space and their symmetries. 

The paper is organized as follows. We first briefly review the relevant tameness results. Then we illustrate the properties of the embedding and the relation with the volume growth in an example. Finally, we discuss the general case and its implications.

\section{Tame Geometry}
\label{sec: tameness}

\subsection{Context and definitions}

Tame sets are a crucial concept to extend notions of finiteness to geometrically continuous objects. The study of their properties helps to bridge the gap between algebraic and analytic geometry. In this section, we briefly present the necessary definitions and results from tame geometry to accurately state the connection between compactifiability and tameness. 

In mathematical terms, a tame set is an element of an o-minimal structure. Let us briefly sketch the definition; for a more in-depth review of the topic we refer the reader to \cite{van1998tame,yomdin2004tame} as well as the introductory summaries found in \cite{Douglas:2022ynw,Douglas:2023fcg,Grimm:2023xqy}.

A \textit{structure} is a collection of sets $\mathcal{S}=(\mathcal{S}_n)_{n\in\mathbb{N}}$, where each $\mathcal{S}_n$ consists of subsets of Euclidean space $\bbR^n$, satisfying the following properties:  $\mathcal{S}$ is closed under Cartesian products and linear projections; 
each $\mathcal{S}_n$ is closed under unions, intersections, and complements; and each $\mathcal{S}_n$ contains the zero loci of all polynomials in $n$ real variables. Sets belonging to a structure are said to be \textit{definable}. This term is chosen to emphasize that they can be constructed by means of set-theoretic operations and, on a more fundamental level, that any statement about such sets can be formulated through a finite number of elementary logical steps. The notion of definability applies to functions as well; a function $f:A\rightarrow  B$ is definable if its graph is a definable subset of $A\times B$.

A structure is \textit{o-minimal} if it satisfies one additional axiom that extends the notion of finiteness from logic theory to geometry:
\begin{itemize}
    \item[--] The definable sets in $\mathcal{S}_1$ are unions of finitely many points and intervals.
\end{itemize}
This axiom places enormous restrictions on the geometry of definable sets, and for this reason sets definable in an o-minimal structure are called tame.

Loosely speaking, a tame set is a geometrical object that has a finite geometric complexity, and therefore cannot have an infinitely discrete property, such as the number of connected components, wrappings, extremal points, or ranks of (co)homology groups. O-minimality is thus the perfect framework in which to answer questions about finiteness, but its axioms are too open to provide specific numerical bounds on geometrically relevant quantities. To address this weakness, a refinement of o-minimality, known as sharp o-minimality, has been developed in the recent years \cite{binyamini2022sharply,binyamini2023tameness}. This class of structures aims to make the idea of finite geometric complexity precise, by introducing a measure of complexity for tame sets. 
    
Sharply o-minimal structures are a subclass of o-minimal structures endowed with a filtration of their definable sets in terms of two natural numbers, $F,D$, named format and degree respectively, which encode the amount of information required to characterize basic geometric features of those sets. In summary, this means that one can group the definable sets in collections $\Omega_{F,D}$, with $\Omega_{F,D}\subseteq \Omega_{F+1,D}$ and $\Omega_{F,D}\subseteq \Omega_{F,D+1}$, that are compatible with the axioms and operations of standard o-minimal structures. In particular, the zero locus of a polynomial of degree $d$ in $n$ variables always belongs to $\Omega_{n, d}$ and if $A_i\in \Omega_{F_i,D_i}$ for $i=1,\dots,k$, then $\bigcup_i A_i,\bigcap_i A_i\in \Omega_{F,D}$ with $F=\max\{F_i\}$ and $D=\sum D_i$. The most important and useful change of sharply o-minimal structures comes from the refinement of the o-minimality axiom itself, setting universal bounds on the number of connected components in terms of the $FD$-filtration:
    \begin{itemize}
        \item[--] There exists a universally fixed function $F\rightarrow P_F$ such that for every $F$, $P_F$ is a polynomial with positive coefficients satisfying that if $A\in \Omega_{F,D}$ with $A\subseteq \mathbb{R}$, then $A$ has at most $P_F(D)$ connected components. 
    \end{itemize}

    This axiom has been shown to be enough to guarantee the existence of similar bounds for higher dimensional sets: 
    \begin{prop}[\cite{binyamini2022sharply}]
    \label{prop: connected components}
        Let $X\subset \mathbb{R}^n$ be a set of format $F$ and degree $D$. Then the number of connected components of $X$ is bounded by a polynomial ${\rm poly}_F(D)$, given by a (possibly n-dependent) universal function $F\rightarrow {\rm poly}_F$.
    \end{prop}
    
    The format and the degree provide a fundamental description of the geometrical/information complexity of the sets in $\Omega_{F,D}$ which addresses a broad range of properties, from topological (constraints on the Betti numbers of a manifold) to algebraic (bounds on the number of solutions of an equation). We refer to \cite{Grimm:2023xqy,Grimm:2024elq} for a deeper exploration of the subject.

    \subsection{Volume bounds}
    
    In order to relate o-minimality to the compactifiability of moduli spaces, we will keep our focus on the number of connected components. For any tame set $A\subset \mathbb{R}^n$ (not necessarily sharply o-minimal) there exists an integer number $b$ such that for any $(n-l)$-dimensional affine plane, the number of connected components of $A\cap P$ is bounded by $b$ for any $0\leq l\leq n$ \cite{yomdin2004tame}. This bound, known as the Gabrielov property \cite{gabrielov1968projections}, has strong implications for the volume growth of an $l$-dimensional tame set $A$ inside an $n$-dimensional $B^n(r)$ as the radius radius $r$ increases. Intuitively, the Gabrielov bound constrains how wrapped the set $A$ can be inside of $B^n(r)$. This guarantees that the volume of $A$ will be proportional to the Euclidean scaling $r^l$. The precise proportionally factor will depend on the wrapping number, which is measured through the number of connected components of the sets $A\cap P$. More formally we have:
    \begin{thr}[\cite{yomdin2004tame}]
    \label{thr: yomdin}
        Let $A\subset \mathbb{R}^n$ be a tame set of dimension $l$. Then for any $n$-dimensional ball $B^n(r)$ in $\mathbb{R}^n$,
        \begin{equation}
            {\rm Vol}_l(A\cap B^n(r))\leq  c(n,l)\, b_{0,n-l}(A)\cdot r^l\,,
            \label{eq: Yomdin bound}
        \end{equation}
        where $b_{0,n-l}(A)$ is the uniform bound of the number of connected components of $A\cap P$ for any $(n-l)$-plane of $\mathbb{R}^n$, $P$, and $c(n,l)$ is a normalization constant given by
        \begin{equation}
         c(n,l)={\rm Vol}_l(B^l(1))\cdot\frac{\Gamma\left(\frac{1}{2}\right)\Gamma\left(\frac{n+1}{2}\right)}{\Gamma\left(\frac{l+1}{2}\right)\Gamma\left(\frac{n-l+1}{2}\right)}\,,
        \end{equation}
        with ${\rm Vol}_l(B^l(1))$ the Euclidean volume of the $l$-dimensional ball of radius one.
    \end{thr}
    The bound \eqref{eq: Yomdin bound} is very reminiscent of the compactifiability condition \eqref{eq: compacty condition}, but there is a key difference that clouds the discussion: while theorem \ref{thr: yomdin} works with a subset of an Euclidean space with a flat metric, the compactifiability condition is focused on the intrinsic properties of the moduli space manifold and its  (non-flat) metric. This problem can be addressed by building an isometric embedding of the moduli space into a higher dimensional Euclidean space, in which theorem \ref{thr: yomdin} can be applied. Such an embedding is guaranteed to exist by Nash theorem \cite{nash1956imbedding}, but, as we will see in the following sections, the tame nature of the manifold might not be preserved under the embedding map.

\section{An example on compactifiability and tameness}

The working example that we will use to illustrate the deep connection between tameness and compactifiability is the hyperbolic plane $\mathbb{H}$. It is a Riemannian manifold consisting of the points $\tau =x+iy\in \mathbb{C}$ with $y>0$ and metric given by
\begin{equation}
    ds^2=\frac{dx^2+dy^2}{y^2}\,. 
    \label{eq: metric H}
\end{equation}
In the following we will see how compactifiability and tameness conditions apply to this example and relate to each other.

\subsection{Hyperbolic plane and compactifiability}

For a generic point $\tau_0=x_0+iy_0$, the associated effective moduli space $\mathcal{M}_D(\phi_0)$ will be
\begin{equation}
\begin{aligned}
         & \mathcal{M}_\mathcal{D}(\tau_0)=\\
  &\{\tau\in \mathbb{H}\ | \ (x-x_0)^2+(y-y_0{\rm cosh}(\mathcal{D}))^2={\rm sinh}^2(\mathcal{D})y_0^2\}\,.
\end{aligned}
\end{equation}
This space is an Euclidean disk in $\mathbb{H}$ of radius $y_0 \sinh(\mathcal{D})$ and center $x_0+i y_0 \cosh(\mathcal{D})$. Using the hyperbolic metric \eqref{eq: metric H}, one then finds
\begin{equation}
    {\rm Vol}(\mathcal{M}_\mathcal{D}) =2\pi(\cosh(\mathcal{D})-1) \,,
    \label{eq: vol MD growth}
\end{equation}
so the asymptotic growth for large geodesic distance is exponential instead of quadratic. We therefore conclude that the hyperbolic plane $\mathbb{H}$ is not compactifiable, as already observed in \cite{Delgado:2024skw}.

However, when one considers the moduli space obtained by quotienting the hyperbolic plane with the action of the standard duality group ${\rm SL}(2,\mathbb{Z})$, the volume becomes finite and thus the compactifiability condition \eqref{eq: compacty condition} is trivially satisfied. This stark contrast highlights the important interplay between the geometry of moduli spaces and the duality groups of effective theories compatible with quantum gravity \cite{Delgado:2024skw}.

Both the upper half-plane $\mathbb{H}$ and the metric function \eqref{eq: metric H} are tame in the simplest o-minimal structure $\mathbb{R}_{\rm alg}$ (see e.g.~\cite{Grimm:2023xqy}). Of course, their restriction to the fundamental domain of the ${\rm SL}(2,\mathbb{Z})$ action is also tame. Consequently, one could naively think that tameness arguments will not be able to distinguish between both cases. However, we must remember that the most powerful result  at our disposal concerning volume growth, theorem \ref{thr: yomdin}, requires a tame isometric embedding of the moduli space into Euclidean space. As we already mentioned in the previous section and as we will explore in detail below, the tame embedding condition is stronger than requiring tameness of the starting manifold. In fact, we will see that it is strong enough to recover the compactifiablity condition \eqref{eq: compacty condition} and extend it beyond the asymptotic regime.

\subsection{Embeddings and tameness in the hyperbolic plane}\label{sec:hyperb}

The Nash embedding theorem \cite{nash1956imbedding} ensures the existence of an isometric embedding of the full hyperbolic plane (or any other Riemannian manifold) into a Euclidean space $\mathbb{R}^n$ for large enough $n$. An explicit, highly non-trivial, realization of such an embedding was found by Blanu{\v{s}}a in \cite{blanuvsa1955einbettung}, requiring $n=6$ and involved complicated non-elementary functions. We present the details in appendix~\ref{ap: H embedding into n=6}. The main observation is that the functions employed in the embedding are periodic in an unbounded domain and thus cannot be tame in any o-minimal structure. 

The situation changes drastically when one restricts the moduli space to the fundamental domain of ${\rm SL}(2,\mathbb{Z})$ defined by 
\begin{equation}
\begin{aligned}
        \mathcal{F}_{{\rm SL}(2,\mathbb{Z})}=&\{x+i y\in \mathbb{C}\ | \ -1/2\leq x \leq 0 \,, \,  x^2+y^2\geq 1\} \ \cup \\
        & \{x+i y\in \mathbb{C}\ | \ 0< x < 1/2 \,, \, x^2+y^2>1\}\,.
        \label{eq: fundamental domain}
\end{aligned}
\end{equation}
In this case, there exists a tame isometric embedding into $\mathbb{R}^3$ given by the section of a pseudosphere \cite{bonahon2009low}. The construction is closely related to the embedding of the Siegel sets of the hyperbolic plane. We refer to appendix \ref{ap: F embedding into n=3} for more details.

Since the fundamental domain $\mathcal{F}_{{\rm SL}(2,\mathbb{Z})}$ admits a tame isometric embedding into $\mathbb{R}^3$, we are in condition to apply theorem \ref{thr: yomdin} to bound the scaling of the volume. First of all, it is immediate to evaluate $c(3,2)=2\pi$. Determining the Gabrielov number is generally more tricky, but in this simple case one can easily deduce it from figure \ref{fig: embeded fundamental domain of H}, from which it is clear that the maximum number of connected components of the intersection between the image of the fundamental domain a straight line in $\mathbb{R}^3$ is bounded by $2$. Therefore, the theorem predicts a polynomial bound on the scaling of the volume 
\begin{equation}
    {\rm Vol}_2(\mathcal{F}^{\rm emb}_{{\rm SL}(2,\mathbb{Z})}\cap B^3(r))\leq  4\pi r^2\,.
    \label{eq: F SL2Z volume bound}
\end{equation}

It is important to note that the fundamental domain is a subset of $\mathbb{H}$ that does not implement the identification between points of the boundary given by the action of ${\rm SL}(2,\mathbb{Z})$. One may then wonder if a tame isometric embedding of the actual moduli space $\mathbb{H}/{\rm SL}(2,\mathbb{Z})$ also exists. Though the explicit construction is much more involved in practice, there are several important results that point towards an affirmative answer. 

First, we recall that the quotient  $\mathbb{H}/{\rm SL}(2,\mathbb{Z})$ is topologically a punctured Riemann sphere $\mathbb{P}^1\backslash \{\infty\}$. The map from $\mathcal{F}_{{\rm SL}(2,\mathbb{Z})}$ to $\mathbb{P}^1\backslash \{\infty\}$ that identifies the points on the boundary of the fundamental domain through the ${\rm SL}(2,\mathbb{Z})$-action is given by the Hauptmodule of the duality group. This map consists of an algebraic combination of powers of the $j$-invariant function. The $j$-invariant function is tame over $\mathcal{F}_{{\rm SL}(2,\mathbb{Z})}$ \cite{peterzil2004uniform}, where it is in fact injective (note that tameness is lost when its domain is extended to the full hyperbolic plane due to the existence of infinite preimages for each point in $\mathbb{P}^1\backslash \{\infty\}$). Consequently, the Hauptmodule map from $\mathcal{F}_{{\rm SL}(2,\mathbb{Z})}$ to $\mathbb{H}/{\rm SL}(2,\mathbb{Z})$ will also be tame. 

Second, it is possible to check that the hyperbolic metric of the upper half-plane is mapped to the Weil-Petersson metric of the modular curve $\mathbb{H}/{\rm SL}(2,\mathbb{Z})$ under the action of the Hauptmodule. One then removes two additional singular points, namely the elliptic points $\tau= -1/2+ \sqrt{3}/2 i$ and $\tau= i$ \footnote{Let $\Gamma$ be a discrete subgroup of $SL(2,\mathbb{R})$. A point $\tau\in \mathbb{H}$ is an elliptic point if it is a fixed point of an element $\gamma\in \Gamma$ with $|{\rm Tr}(\gamma)|<2$.}, and works on $\mathbb{P}^1\backslash \{0,1,\infty\}$. The metric on this space is derived from a Kähler potential that can be expanded as a function of the complex structure modulus in terms of the two independent periods of the modular curve
\begin{equation}
    K_{\rm cs}=-\log\left[i (\Pi^0(z)\overline{\Pi}_0(\bar{z})-\overline{\Pi}^0(\bar{z})\Pi_0(z))\right]\,.
\end{equation}
 A recent result from o-minimality proves the tameness of the periods \cite{bakker2020tame}, which means that the Weil-Petersson metric obtained from $g_{z\bar{z}}=\partial_z\partial_{\bar{z}} K_{\rm cs}$ will also be tame. 

The discussion above recontextualizes the role of the duality group and the importance of quotienting the moduli space by its action. The identification of points in the same orbit keeps the $j$-function and the period map in a domain where they are tame. In the particular case of the hyperbolic plane, as in many others, this requirement results in a moduli space of finite volume.

Having established the tame nature of $\mathbb{H}/{\rm SL}(2,\mathbb{Z})$ and its metric, the only potential obstruction to the existence of a tame isometric embedding into Euclidean space are the three special limit points of the fundamental domain $\mathcal{F}_{{\rm SL}(2,\mathbb{Z})}$. One of them ($\tau\rightarrow i\infty$) gives rise to the cusp of the punctured sphere, while the two others correspond to elliptic points previously mentioned. Out of the three, the most problematic is the cusp, which is the only one that describes a point at infinite distance. The tame embeddability of this geometric feature into Euclidean space is shown using the notion of Siegel sets in appendix \ref{ap: F embedding into n=3}. Consequently, we expect $\mathbb{H}/{\rm SL}(2,\mathbb{Z})$ to admit an isometric tame embedding and therefore obey similar bounds to \eqref{eq: F SL2Z volume bound}.

What is then the difference between the spaces $\mathbb{H}$ and $\mathbb{H}/{\rm SL}(2,\mathbb{Z})$ with regards to the isometric embedding? The hyperbolic plane is simply too large to  be embedded in Euclidean space without folding it infinitely many times. In this sense, the negative curvature, which controls the folding, presents an obstruction to the tame embedding. The length of any bounded horizontal segment $x\in (-c,c)$ diverges when it approaches $y=0$ and such growth cannot be accounted for by any tame embedding.  This untamable behavior appears when approaching regions at infinite distance in moduli space. In \cite{fontenele2014complexity}, the obstruction is more rigorously formalized and extended to a certain class of simply connected Riemannian manifolds with negative curvature. The moduli space $\mathbb{H}/{\rm SL}(2,\mathbb{Z})$ is not simply connected, so it evades the premise of the theorem. Moreover, it only has one limit point at infinite distance and so the issue regarding the divergence of segments does not arise. More generally, we expect any duality group acting over $\mathbb{H}$ whose fundamental domain has a finite number of infinite distance points to be tamely embeddable. This will include Fuchsian groups of the first kind, such as the congruence subgroups of ${\rm SL}(2,\mathbb{Z})$ \cite{voight2021quaternion, Delgado:2024skw}.

In the following section we will show that this picture is very general and can be easily pushed back to provide a bound on the scaling of the volume of the original manifold with respect to the geodesic distance. We will also explore how the coefficient in front of the polynomial growth can be related to the complexity of the manifold and its embedding.

\section{General picture}

\subsection{Volume Growth}
\label{subsec: volume growth}

Now that we have understood the importance of the embedding and the subtle but crucial distinction between a tame manifold and a tame embedding into Euclidean space, let us present the general result that relates this framework to the compactifiability conjecture. 

We start by introducing some notation. We denote by $\mathcal{M}$ the candidate moduli space under consideration (either the complete space, its fundamental domain under some duality group or the quotient space under said duality) and by $g$ the metric in that space. Let $\phi$ be the isometrical embedding of the moduli space into an $N-$dimensional Euclidean space, that is $\phi: \mathcal{M}\rightarrow \mathbb{R}^N$ with $g=\phi^*\eta$, where $\eta$ is the $N-$dimensional Euclidean metric. Furthermore, we define $\mathcal{M}^{\rm emb}(x_0)=\phi(\mathcal{M}(x_0))$ to be the image under the embedding of the the moduli space with a marked point $x_0$. The manifold $\mathcal{M}_\mathcal{D}^{\rm emb}(x_0)$ will be the submanifold of $\mathcal{M}^{\rm emb}(x_0)$ with all points that are maximally a geodesic distance $\mathcal{D}$ away from $x_0$. Finally, let us introduce $\mathcal{M}^{\rm emb}(x_0) \cap B^{\rm emb}(r)$, where $B^{\rm emb}(r)$ is a $N$-dimensional ball of radius $r$ centered around $\phi(x_0)$ measured in the Euclidean distance. The Euclidean geometry of the embedding space implies 
\begin{equation}
  \mathcal{M}_\mathcal{D}^{\rm emb}(x_0) \subset \mathcal{M}^{\rm emb}(x_0) \cap B^{\rm emb}(r)\ \ \text{for}\ \ \mathcal{D}\leq r\,.
  \label{eq: Md subset condition}
\end{equation}
Since the isometric embedding preserves the volumes, $ {\rm Vol}(\mathcal{M}^{\rm emb}_\mathcal{D}(x_0))= {\rm Vol}(\mathcal{M}_\mathcal{D}(x_0))$,  it trivially follows that if $r\geq \mathcal{D}$ then
\begin{equation}\label{eq:VolDvsBemb}
   {\rm Vol}(\mathcal{M}_\mathcal{D}(x_0))\leq {\rm Vol}(\mathcal{M}^{\rm emb}(x_0) \cap B^{\rm emb}(r))\,.
\end{equation}
Furthermore, theorem \ref{thr: yomdin} tells us that if the embedded moduli space $\mathcal{M}^{\rm emb}(x_0)$ is a tame set, it satisfies the following bound
\begin{equation}
    {\rm Vol}(\mathcal{M}^{\rm emb}(x_0)\cap B^{\rm emb}(r))\leq C(\mathcal{M}^{\rm emb}) r^l\,,
        \label{eq: tameness vol scaling}
\end{equation}
with $l={\rm dim}(\mathcal{M})$ and $C(\mathcal{M}^{\rm emb})$ a global coefficient depending on the complexity of the embedding. Choosing a ball of radius $r=\mathcal{D}$ we can simply combine \eqref{eq: Md subset condition}  and \eqref{eq: tameness vol scaling} to conclude

\begin{equation}
   \hspace{-0.5cm} \boxed{{\rm Vol}(\mathcal{M}_\mathcal{D}(x_0))\leq  C(\mathcal{M}^{\rm emb}) \mathcal{D}^l\,,}
    \label{eq: THE BOUND}
\end{equation}
which shows that if the moduli space $\mathcal{M}$ is a tame Riemannian manifold admitting a tame embedding into Euclidean space, the compactifiability conjecture \eqref{eq: compacty condition} is automatically satisfied. 

Conversely, result \eqref{eq: THE BOUND} can be used to rule out the existence of a tame isometric embedding. This is the case of the hyperbolic plane $\mathbb{H}$: the volume grows exponentially in the asymptotic limit (see \eqref{eq: vol MD growth}), which is incompatible with any tame embedding. This formalizes the observation of the previous section regarding the non-tame nature of Blanu{\v{s}}a's embedding and extends it from a particular property of that single example to a general result applying to any other potential construction. 

It is also worth noting that the bound provided by \eqref{eq: THE BOUND}, despite depending on the global properties of the embedding, is a local constraint that holds for any point $x_0$ and any value of the geodesic distance $\mathcal{D}$. In this sense, it provides a more powerful statement about the volume growth than the original compactifiability condition \eqref{eq: compacty condition}, which only applied in the asymptotic limit $\mathcal{D}\rightarrow \infty$.

Finally, we note that any tame embedding for which equation \eqref{eq:VolDvsBemb} holds already implies the compactifiability condition. Therefore, the requirement that the embedding is isometric can be weakened slightly. However, this weakening is rather subtle. An embedding which contracts distances implies equation \eqref{eq: Md subset condition} but not that $\text{Vol}(\cM_\cD)\leq \text{Vol}(\cM_\cD^{\text{emb}})$, whereas an embedding which expands distances implies that $\text{Vol}(\cM_\cD)\leq \text{Vol}(\cM_\cD^{\text{emb}})$ but not that \eqref{eq: Md subset condition} holds. The embedding must be such that the rescaling of distances compensates the wrapping of the embedded manifold in the Euclidean target space, in such a way that  equation \eqref{eq:VolDvsBemb} is satisfied. An isometric embedding automatically achieves this, which makes it the natural notion to consider.

\subsection{Complexity}

In section \ref{sec: tameness} we introduced a refinement of o-minimality, sharp o-minimality, that naturally has a notion of complexity characterized by a pair of numbers: the format $F$ and the degree $D$. If the embedding is not only tame, but definable in a sharply o-minimal structure, it is possible to go beyond purely finiteness statements and set explicit bounds on the coefficients of the volume scaling. In particular, it is possible to bound the Gabrielov numbers $b_{0,n-l}(A)$ for $A\in\Omega_{F, D}$.

From the proposition \ref{prop: connected components}, we know that in a sharply o-minimal structure there exists a universal function ${\rm poly}_{F}(D)$ that bounds the number of connected components of any  definable set $A\subset \mathbb{R}^n$ in $\Omega_{F, D}$. Let us also recall that for a given  $l$-dimensional set $A\subset \mathbb{R}^n$, the Gabrielov number $b_{0,n-l}(A)$ is given by the maximum number of connected components of $A\cap P$ for any $(n-l)$-dimensional affine plane $P$. The family of affine planes is definable in any sharply o-minimal structure and satisfies $P\in \Omega_{n, l}$.  Let us assume that  $\mathcal{M}^{\rm emb}\in \Omega_{F, D}$. Then, from the axioms of sharp o-minimality, $\mathcal{M}^{\rm emb}\cap P$ is definable in the same structure and satisfies $\mathcal{M}^{\rm emb}\cap P\in \Omega_{F,(D+l)}$. We thus have
\begin{equation}
    b_{0,n-l}(\mathcal{M}^{\rm emb})\leq {\rm poly}_{F}(D+l)\,.
\end{equation}

Then, we can refine \eqref{eq: THE BOUND} for the moduli spaces admitting a tame embedding in a sharply o-minimal structure with $\mathcal{M}^{\rm emb}\in \Omega_{F,D}$:
\begin{equation}
    C(\mathcal{M}^{\rm emb})\leq c(n,l)\cdot {\rm poly}_{F}(D+l)\,.
    \label{eq: complexity bound}
\end{equation}

Note this bounding function ${\rm poly}_{F}(D+l)$ is universal for sets in $\mathbb{R}^n$, with fixed $n$, throughout the sharp o-minimal structure. Therefore, once the embedding dimension $n$ has been established, the dependence on the particular choice of moduli space is present only through its dimensionality $l$ and the pair of numbers $(F,D)$ that characterize the complexity of the isometric Euclidean embedding. 

Relation \eqref{eq: complexity bound} can be used in two different ways. We have just seen how, knowing the complexity of the embedding, one can constrain the coefficient in front of the volume scaling. An equally interesting application of this relation is to use the behavior of the volume scaling with the geodesic distance as a proxy for the complexity of the construction, since it sets a lower bound on the format and degree of the embedding.

\section{\label{sec:conclusion} Outlook}

We have observed that the existence of a tame isometric embedding of the moduli space into flat space is a sufficient condition to recover the compactifiability criterion \eqref{eq: compacty condition}. Following up on the claim of \cite{Delgado:2024skw} that the latter condition is a universal property for quantum gravity theories, we conjecture that any effective field theory compatible with quantum gravity must have a moduli space that admits a tame isometric embedding into Euclidean space. This refines the idea of the original tameness conjecture in \cite{Grimm:2021vpn}, where the tameness of the manifold and the coupling functions was considered. As we have seen in this note, tameness of the moduli space, merely viewed as a manifold with a metric, is not strong enough to recover the volume scaling properties expected from a theory of quantum gravity. Instead we noted the importance of tameness of certain functions on the moduli space when analyzing finiteness. In this work we have focused on the isometric embedding, but we expect that these observations can be recast in terms of other functions relevant in the effective description, such as the periods or the modular functions. This expectation stems from the fact that the described constraints are similarly needed in proving the tameness of the period map for Calabi-Yau moduli spaces. Studying the tame properties of these functions and the relations among them constitutes a large field of research that deserves further consideration.

Reformulating the compactifiability criterion in terms of tameness offers several advantages. First, it ties this condition into a general framework and thereby unifies it with other finiteness statements, e.g.~about the number of extrema of a scalar potential. Second, it provides a sharp local characterization of the volume growth that goes beyond asymptotic statements. Despite its local validity, universal bounds on the coefficients of the scaling are formulated in terms of global topological properties of the moduli space and its embedding. When extending the tameness principle to sharp o-minimality, the coefficients of the volume growth can be recast as functions of the complexity of the moduli space. We thus can establish a quantitative connection to complexity and finiteness of information through the study of the coefficient present in the polynomial growth. Following more generally on the last point, a notion of complexity for an effective field theory \cite{Grimm:2023xqy, Grimm:2024elq} could be used to connect with the distance conjecture through the species scale \cite{Dvali:2007hz, Dvali:2007wp, Dvali:2009ks, Castellano:2021mmx, vandeHeisteeg:2022btw}, which itself provides a cutoff to the effective field theory that keeps the number of light states finite.

We stress that our findings also show a direct link between tame embeddability and the existence of dualities. The example of the hyperbolic plane highlighted that the moduli space without taking the duality quotient is too large to be tamely isometrically embedded into Euclidean space. In fact, in this example both T- and S-duality are required to render the quotiented moduli space small enough to admit a tame embedding, while T- and S-duality individually are not sufficient. In general, recall from section \ref{sec:hyperb} that the main result of \cite{fontenele2014complexity} implies roughly speaking that simply connected negatively curved manifolds cannot admit a tame embedding. Discrete duality quotients break simply connectedness, so that the assumption of the theorem is evaded. In fact, as noted in the context of marked moduli spaces in \cite{Raman:2024fcv},  the breaking of simply connectedness is always a consequence of the existence of dualities, which further establishes the connection between tameness and dualities. Furthermore, the tameness of the isometric embedding is reminiscent of the tameness of the period map on the moduli space, for which the quotient by a sufficiently large duality group is an essential part of the proof \cite{bakker2020tame}. Moreover, the proper consideration of duality quotients is also essential in the finiteness proof of \cite{Bakker:2021uqw}. In the future, it would be interesting to consider more generally what additional properties of the duality groups could be inferred from tame embeddability. In reference \cite{Delgado:2024skw}, the duality groups of algebraically compactifiable moduli spaces are proved to be semisimple. Algebraic compactifiability is stronger than standard compactifiability and seems closely related to tame isometric Euclidean embeddings. The latter has the added advantage that it holds beyond the cases where the moduli space is a complex manifold, implying that tame geometry could further extend these results. These questions are worth exploring in further research.

Stating the precise requirements that ensure that a tame Riemannian manifold admits a tame isometric embedding into flat space is an interesting open problem. From a mathematical perspective this would require to formulate and prove a tame version of the Nash embedding theorem. A useful intermediate result would be to establish the theorem for compact tame manifolds. This holds true for the subclass of compact analytic manifolds. The tameness of these manifolds follows from the tameness of restricted analytic functions~\cite{van1994real}. To establish the statement of tame embeddability, we can then use the analytic Nash embedding theorem~\cite{greene1971analytic}, which guarantees the existence of an analytic isometric embedding for any Riemannian manifold with an analytic metric.
Returning to moduli spaces, it is clear that demanding compactness would be a too strong  condition. In fact, it was conjectured that generally  moduli space should be non-compact in the original formulation of the distance conjecture \cite{Ooguri:2006in}. Nevertheless, from the relation between compactness and tameness of the embedding we infer that infinite distance limits are the main source of potential conflict. Given the general result of \cite{fontenele2014complexity}, spaces with negative curvature and infinite distance boundaries are especially problematic when considering tame embeddings. Consequently, our work suggests a deep connection between the curvature of moduli spaces \cite{Marchesano:2023thx, Marchesano:2024tod, Castellano:2024gwi}, the distance conjecture \cite{vandeHeisteeg:2022btw,Calderon-Infante:2023ler}, and the finiteness of complexity and information \cite{Grimm:2023lrf, Grimm:2024elq}. A characterization of the properties that infinite distance limits must satisfy in order to verify our new isometric embeddability conjecture could further enhance the understanding of all these topics.

\begin{acknowledgments}
We would like to thank Gal Binyamini, Martín Carrascal, Raf Cluckers, Andrei Gabrielov, Damian van de Heisteeg, Pyry Kuusela, Dmitry Novikov, Thorsten Schimannek,  Javier Subils, and Stefan Vandoren for useful discussions and comments. This research is supported, in part, by the Dutch Research Council (NWO) via a Vici grant.
\end{acknowledgments}

\appendix

\section{Isometric embedding of the hyperbolic plane in $\mathbb{R}^6$}
\label{ap: H embedding into n=6}

For completeness sake we provide the embedding into $\mathbb{R}^6$ as described in \cite{blanuvsa1955einbettung} (see also \cite{borisenko2001isometric}). To do so, we will need to introduce a new parametrization of the hyperbolic plane using the Poincaré disk description. This is obtained through the map  $s=-i\frac{1+i\tau}{1-i\tau}$,
which satisfies $|s|\leq 1$ for any $\tau=x+iy\in \mathbb{H}$. We take one additional step to describe every the complex point $s$ in terms of hyperbolic polar coordinates $(u,v)$ of the form
$s=\frac{\sinh(u)}{1+\cosh(u)}e^{iv}$, 
where $u\geq 0$ and $v\in [-\pi,\pi)$. It is in these hyperbolic coordinates in which Blanu{\v{s}}a's embedding is constructed. To start, one needs to auxiliary functions 
\begin{equation}
    \begin{aligned}
    \psi_1(u) =& e^{2 \left( \left[ \frac{|u| + 1}{2} \right] \right) + 5}\,,\qquad \psi_2(u) = e^{2 [|u|/2 ] + 6}\,,\\
    A =& \int_0^1 \sin \pi \xi \, e^{-1/\sin^2 \pi \xi} \, d\xi\,,\\
    \varphi_1(u) =& \left( \frac{1}{A} \int_0^{u+1} \sin \pi \xi \, e^{-1/\sin^2 \pi \xi} \, d\xi \right)^{1/2}\,,\\
    \varphi_2(u) =& \left( \frac{1}{A} \int_0^u \sin \pi \xi \, e^{-1/\sin^2 \pi \xi} \, d\xi \right)^{1/2}\,,\\
    f_1(u) =& \frac{\varphi_1(u)}{\psi_1(u)} \sinh u\,,\qquad f_2(u) = \frac{\varphi_2(u)}{\psi_2(u)} \sinh u\,,
    \end{aligned}
\end{equation}
where $[\cdot]$ stands for the integral part of the bracket expression. Let $x_i$ ($i=1,\dots, 6)$ be the Cartesian coordinates in $\mathbb{R}^6$. The embedding in $\mathbb{R}^6$ of the hyperbolic plane with the line element $ds^2=du^2+{\rm sinh}^2 u dv^2=(dx^2+dy^2)/y^2$ is given by
\begin{align}
    x_1 &= \int_{0}^{u} \sqrt{1 - f_1'^2(\xi) - f_2'^2(\xi)} \, d\xi\,, \qquad 
    x_2 = v\,,     \label{eq: E6 embedding}\\
    x_3 &= f_1(u) \cos(v \, \psi_1(u))\,, \quad
    x_4 = f_1(u) \sin(v \, \psi_1(u))\,, \nonumber\\
    x_5 &= f_2(u) \cos(v \, \psi_2(u))\,, \quad
    x_6 = f_2(u) \sin(v \, \psi_2(u))\,.\nonumber
\end{align}

Apart from the unintuitive nature of  the embedding, it is important to highlight the presence of a trigonometric function $e^{-1/\sin^2\pi u}$ taking values in an unbounded domain $u\geq 0$. This cannot be definable in any o-minimal structure, which leads us to conclude that the embedding is not tame. Note this result does not exclude the existence of an alternative tame isometric embedding of $\mathbb{H}$. Such a scenario is ruled out due to the asymptotic exponential volume growth (see discussion of section \ref{subsec: volume growth}).

\section{Tame isometric embedding of the fundamental domain of ${\rm SL}(2,\mathbb{Z})$}
\label{ap: F embedding into n=3}

 The simplest way to construct an isometric embedding of the fundamental domain $ \mathcal{F}_{{\rm SL}(2,\mathbb{Z})}$ given in \eqref{eq: fundamental domain} is by providing an embedding of a Siegel set. These sets trade the injectivity under the group action in favor of the simplicity of their geometrical shape, while preserving the relevant notions of finiteness that characterize the fundamental domains. Siegel sets are deeply rooted in o-minimal geometry and  have played an important role in proving the tameness of the period maps \cite{bakker2020tame}. Their precise definition is given in \cite{borel2006compactifications}. For our purposes, we can think of them as Euclidean boxes in $\mathbb{H}$ of the form $x+iy\in \mathbb{H}$ with $-c<x<c$ and $y>\lambda$ for $c,\lambda>0$.

\begin{figure}[h!]
    \centering
\includegraphics[width=0.72\linewidth]{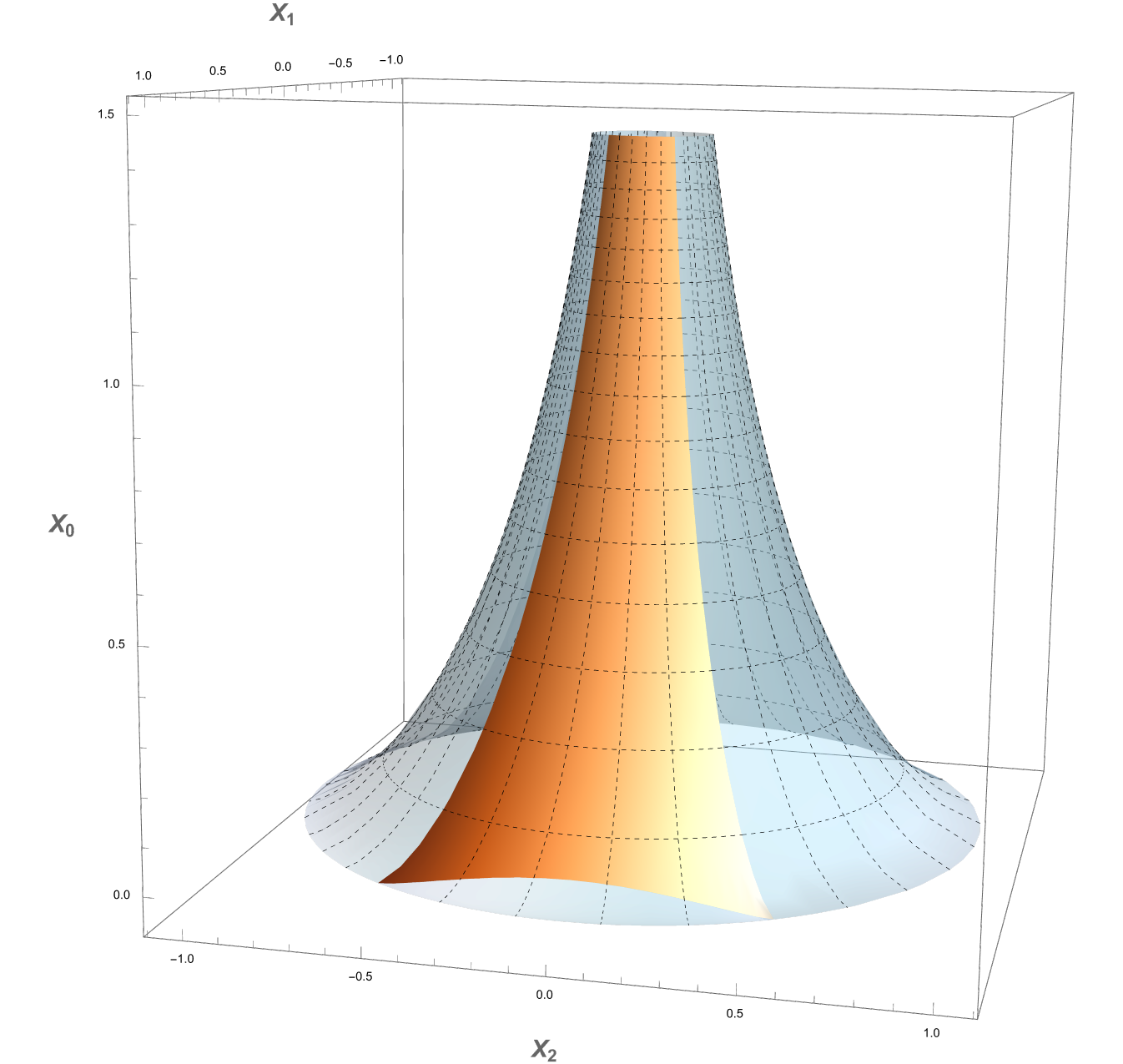}
    \caption{Image of a Siegel set $\mathcal{S}_{\sqrt{3}/2}$ (blue) and the fundamental domain  $\mathcal{F}_{{\rm SL}(2,\mathbb{Z})}$ (copper) under the action of the embedding map \eqref{eq: E3 embedding} into Euclidean space $\mathbb{R}^3$.}
    \label{fig: embeded fundamental domain of H}
\end{figure}
 
 In particular, we will embed the following Siegel set: $\mathcal{S}_c: (x,y)\in \mathbb{H}$ with $-c\, \pi<x<c\, \pi $ and $y>c$, taking $c=\sqrt{3}/2$. Clearly, $\mathcal{S}_{\sqrt{3}/2}$ fully contains $\mathcal{F}_{{\rm SL}(2,\mathbb{Z})}$. The embedding is given as follows \cite{bonahon2009low}
\begin{equation}
    \begin{gathered}
        X_0 = t - \tanh t\,, \qquad
        X_1 = \frac{\sqrt{3}\cos \left( 2 x/\sqrt{3} \right)}{2 y}\,, \\
        X_2 =  \frac{\sqrt{3}\sin \left( 2x/\sqrt{3} \right)}{2 y}\,,
        \label{eq: E3 embedding} 
    \end{gathered}
\end{equation}
with $t=\arccosh(2y/\sqrt{3})$. The images under the embedding map  of the Siegel set and the fundamental domain are depicted in figure \ref{fig: embeded fundamental domain of H}. All the functions involved in the map are tame when restricting the variables $(x,y)$ to the domain $\mathcal{S}_{\sqrt{3}/2}$ and furthermore $dX_0^2+dX_1^2+dX_2^2=\frac{dx^2+dy^2}{y^2}\,$. We conclude that the both the Siegel sets $\mathcal{S}_c$ and the fundamental domain $ \mathcal{F}_{{\rm SL}(2,\mathbb{Z})}$ admit a tame isometric embedding into $\mathbb{R}^3$.

\bibliographystyle{utphys}
\bibliography{biblio}

\providecommand{\href}[2]{#2}\begingroup\raggedright\begin{thebibliography}{10}

\bibitem{Vafa:2005ui}
C.~Vafa, ``{The String landscape and the swampland},'' \href{http://arxiv.org/abs/hep-th/0509212}{{\ttfamily arXiv:hep-th/0509212}}.

\bibitem{Arkani-Hamed:2006emk}
N.~Arkani-Hamed, L.~Motl, A.~Nicolis, and C.~Vafa, ``{The String landscape, black holes and gravity as the weakest force},'' \href{http://dx.doi.org/10.1088/1126-6708/2007/06/060}{{\em JHEP} {\bfseries 06} (2007) 060}, \href{http://arxiv.org/abs/hep-th/0601001}{{\ttfamily arXiv:hep-th/0601001}}.

\bibitem{Banks:2010zn}
T.~Banks and N.~Seiberg, ``{Symmetries and Strings in Field Theory and Gravity},'' \href{http://dx.doi.org/10.1103/PhysRevD.83.084019}{{\em Phys. Rev. D} {\bfseries 83} (2011) 084019}, \href{http://arxiv.org/abs/1011.5120}{{\ttfamily arXiv:1011.5120 [hep-th]}}.

\bibitem{Ooguri:2006in}
H.~Ooguri and C.~Vafa, ``{On the Geometry of the String Landscape and the Swampland},'' \href{http://dx.doi.org/10.1016/j.nuclphysb.2006.10.033}{{\em Nucl. Phys. B} {\bfseries 766} (2007) 21--33}, \href{http://arxiv.org/abs/hep-th/0605264}{{\ttfamily arXiv:hep-th/0605264}}.

\bibitem{Douglas:2003um}
M.~R. Douglas, ``{The Statistics of string / M theory vacua},'' \href{http://dx.doi.org/10.1088/1126-6708/2003/05/046}{{\em JHEP} {\bfseries 05} (2003) 046}, \href{http://arxiv.org/abs/hep-th/0303194}{{\ttfamily arXiv:hep-th/0303194}}.

\bibitem{Acharya:2006zw}
B.~S. Acharya and M.~R. Douglas, ``{A Finite landscape?},'' \href{http://arxiv.org/abs/hep-th/0606212}{{\ttfamily arXiv:hep-th/0606212}}.

\bibitem{Hamada:2021yxy}
Y.~Hamada, M.~Montero, C.~Vafa, and I.~Valenzuela, ``{Finiteness and the swampland},'' \href{http://dx.doi.org/10.1088/1751-8121/ac6404}{{\em J. Phys. A} {\bfseries 55} no.~22, (2022) 224005}, \href{http://arxiv.org/abs/2111.00015}{{\ttfamily arXiv:2111.00015 [hep-th]}}.

\bibitem{Grimm:2021vpn}
T.~W. Grimm, ``{Taming the landscape of effective theories},'' \href{http://dx.doi.org/10.1007/JHEP11(2022)003}{{\em JHEP} {\bfseries 11} (2022) 003}, \href{http://arxiv.org/abs/2112.08383}{{\ttfamily arXiv:2112.08383 [hep-th]}}.

\bibitem{Douglas:2022ynw}
M.~R. Douglas, T.~W. Grimm, and L.~Schlechter, ``{The Tameness of Quantum Field Theory, Part I -- Amplitudes},'' \href{http://arxiv.org/abs/2210.10057}{{\ttfamily arXiv:2210.10057 [hep-th]}}.

\bibitem{Douglas:2023fcg}
M.~R. Douglas, T.~W. Grimm, and L.~Schlechter, ``{The Tameness of Quantum Field Theory, Part II -- Structures and CFTs},'' \href{http://arxiv.org/abs/2302.04275}{{\ttfamily arXiv:2302.04275 [hep-th]}}.

\bibitem{Bakker:2021uqw}
B.~Bakker, T.~W. Grimm, C.~Schnell, and J.~Tsimerman, ``{Finiteness for self-dual classes in integral variations of Hodge structure},'' \href{http://arxiv.org/abs/2112.06995}{{\ttfamily arXiv:2112.06995 [math.AG]}}.

\bibitem{Grimm:2023lrf}
T.~W. Grimm and J.~Monnee, ``{Finiteness theorems and counting conjectures for the flux landscape},'' \href{http://dx.doi.org/10.1007/JHEP08(2024)039}{{\em JHEP} {\bfseries 08} (2024) 039}, \href{http://arxiv.org/abs/2311.09295}{{\ttfamily arXiv:2311.09295 [hep-th]}}.

\bibitem{Grimm:2022sbl}
T.~W. Grimm, S.~Lanza, and C.~Li, ``{Tameness, Strings, and the Distance Conjecture},'' \href{http://dx.doi.org/10.1007/JHEP09(2022)149}{{\em JHEP} {\bfseries 09} (2022) 149}, \href{http://arxiv.org/abs/2206.00697}{{\ttfamily arXiv:2206.00697 [hep-th]}}.

\bibitem{Grimm:2023xqy}
T.~W. Grimm, L.~Schlechter, and M.~van Vliet, ``{Complexity in tame quantum theories},'' \href{http://dx.doi.org/10.1007/JHEP05(2024)001}{{\em JHEP} {\bfseries 05} (2024) 001}, \href{http://arxiv.org/abs/2310.01484}{{\ttfamily arXiv:2310.01484 [hep-th]}}.

\bibitem{Grimm:2024elq}
T.~W. Grimm and M.~van Vliet, ``{On the Complexity of Quantum Field Theory},'' \href{http://arxiv.org/abs/2410.23338}{{\ttfamily arXiv:2410.23338 [hep-th]}}.

\bibitem{Delgado:2024skw}
M.~Delgado, D.~van~de Heisteeg, S.~Raman, E.~Torres, C.~Vafa, and K.~Xu, ``{Finiteness and the Emergence of Dualities},'' \href{http://arxiv.org/abs/2412.03640}{{\ttfamily arXiv:2412.03640 [hep-th]}}.

\bibitem{Lu:2005bj}
Z.~Lu and X.~Sun, ``{On the Weil-Petersson volume and the first Chern class of the moduli space of Calabi-Yau manifolds},'' \href{http://dx.doi.org/10.1007/s00220-005-1441-3}{{\em Commun. Math. Phys.} {\bfseries 261} (2006) 297--322}, \href{http://arxiv.org/abs/math/0510021}{{\ttfamily arXiv:math/0510021}}.

\bibitem{van1998tame}
L.~Van~den Dries, {\em Tame topology and o-minimal structures}, vol.~248.
\newblock Cambridge university press, 1998.

\bibitem{yomdin2004tame}
Y.~Yomdin and G.~Comte, {\em Tame geometry with application in smooth analysis}.
\newblock Springer, 2004.

\bibitem{binyamini2022sharply}
G.~Binyamini, D.~Novikov, and B.~Zack, ``Sharply o-minimal structures and sharp cellular decomposition,'' {\em arXiv preprint arXiv:2209.10972} (2022) .

\bibitem{binyamini2023tameness}
G.~Binyamini and D.~Novikov, ``Tameness in geometry and arithmetic: beyond o-minimality,'' in {\em International congress of mathematicians}, pp.~1440--1461.
\newblock 2023.

\bibitem{gabrielov1968projections}
A.~M. Gabrielov, ``Projections of semi-analytic sets,'' {\em Functional Analysis and its applications} {\bfseries 2} no.~4, (1968) 282--291.

\bibitem{nash1956imbedding}
J.~Nash, ``The imbedding problem for {R}iemannian manifolds,'' {\em Annals of mathematics} {\bfseries 63} no.~1, (1956) 20--63.

\bibitem{blanuvsa1955einbettung}
D.~Blanu{\v{s}}a, ``{\"U}ber die {E}inbettung hyperbolischer {R}{\"a}ume in euklidische {R}{\"a}ume,'' {\em Monatshefte f{\"u}r Mathematik} {\bfseries 59} no.~3, (1955) 217--229.

\bibitem{bonahon2009low}
F.~Bonahon, {\em Low-dimensional geometry: From Euclidean surfaces to hyperbolic knots}, vol.~49.
\newblock American Mathematical Soc., 2009.

\bibitem{peterzil2004uniform}
Y.~Peterzil and S.~Starchenko, ``Uniform definability of the {W}eierstrass $\wp$ functions and generalized tori of dimension one,'' {\em Selecta Math.(NS)} {\bfseries 10} no.~4, (2004) 525--550.

\bibitem{Note1}
Let $\Gamma $ be a discrete subgroup of $SL(2,\protect \mathbb {R})$. A point $\tau \in \protect \mathbb {H}$ is an elliptic point if it is a fixed point of an element $\gamma \in \Gamma $ with $|{\protect \rm Tr}(\gamma )|<2$.

\bibitem{bakker2020tame}
B.~Bakker, B.~Klingler, and J.~Tsimerman, ``Tame topology of arithmetic quotients and algebraicity of hodge loci,'' {\em Journal of the American Mathematical Society} {\bfseries 33} no.~4, (2020) 917--939.

\bibitem{fontenele2014complexity}
F.~Fontenele and F.~Xavier, ``On the complexity of isometric immersions of hyperbolic spaces in any codimension,'' {\em arXiv preprint arXiv:1410.8465} (2014) .

\bibitem{voight2021quaternion}
J.~Voight, {\em Quaternion algebras}.
\newblock Springer Nature, 2021.

\bibitem{Dvali:2007hz}
G.~Dvali, ``{Black Holes and Large N Species Solution to the Hierarchy Problem},'' \href{http://dx.doi.org/10.1002/prop.201000009}{{\em Fortsch. Phys.} {\bfseries 58} (2010) 528--536}, \href{http://arxiv.org/abs/0706.2050}{{\ttfamily arXiv:0706.2050 [hep-th]}}.

\bibitem{Dvali:2007wp}
G.~Dvali and M.~Redi, ``{Black Hole Bound on the Number of Species and Quantum Gravity at LHC},'' \href{http://dx.doi.org/10.1103/PhysRevD.77.045027}{{\em Phys. Rev. D} {\bfseries 77} (2008) 045027}, \href{http://arxiv.org/abs/0710.4344}{{\ttfamily arXiv:0710.4344 [hep-th]}}.

\bibitem{Dvali:2009ks}
G.~Dvali and D.~Lust, ``{Evaporation of Microscopic Black Holes in String Theory and the Bound on Species},'' \href{http://dx.doi.org/10.1002/prop.201000008}{{\em Fortsch. Phys.} {\bfseries 58} (2010) 505--527}, \href{http://arxiv.org/abs/0912.3167}{{\ttfamily arXiv:0912.3167 [hep-th]}}.

\bibitem{Castellano:2021mmx}
A.~Castellano, A.~Herr\'aez, and L.~E. Ib\'a\~nez, ``{IR/UV mixing, towers of species and swampland conjectures},'' \href{http://dx.doi.org/10.1007/JHEP08(2022)217}{{\em JHEP} {\bfseries 08} (2022) 217}, \href{http://arxiv.org/abs/2112.10796}{{\ttfamily arXiv:2112.10796 [hep-th]}}.

\bibitem{vandeHeisteeg:2022btw}
D.~van~de Heisteeg, C.~Vafa, M.~Wiesner, and D.~H. Wu, ``{Moduli-dependent species scale},'' \href{http://dx.doi.org/10.4310/bpam.2024.v1.n1.a1}{{\em Beijing J. Pure Appl. Math.} {\bfseries 1} no.~1, (2024) 1--41}, \href{http://arxiv.org/abs/2212.06841}{{\ttfamily arXiv:2212.06841 [hep-th]}}.

\bibitem{Raman:2024fcv}
S.~Raman and C.~Vafa, ``{Swampland and the Geometry of Marked Moduli Spaces},'' \href{http://arxiv.org/abs/2405.11611}{{\ttfamily arXiv:2405.11611 [hep-th]}}.

\bibitem{van1994real}
L.~van~den Dries and C.~Miller, ``On the real exponential field with restricted analytic functions,'' {\em Israel Journal of Mathematics} {\bfseries 85} (1994) 19--56.

\bibitem{greene1971analytic}
R.~E. Greene and H.~Jacobowitz, ``Analytic isometric embeddings,'' {\em Annals of Mathematics} {\bfseries 93} no.~1, (1971) 189--204.

\bibitem{Marchesano:2023thx}
F.~Marchesano, L.~Melotti, and L.~Paoloni, ``{On the moduli space curvature at infinity},'' \href{http://dx.doi.org/10.1007/JHEP02(2024)103}{{\em JHEP} {\bfseries 02} (2024) 103}, \href{http://arxiv.org/abs/2311.07979}{{\ttfamily arXiv:2311.07979 [hep-th]}}.

\bibitem{Marchesano:2024tod}
F.~Marchesano, L.~Melotti, and M.~Wiesner, ``{Asymptotic curvature divergences and non-gravitational theories},'' \href{http://arxiv.org/abs/2409.02991}{{\ttfamily arXiv:2409.02991 [hep-th]}}.

\bibitem{Castellano:2024gwi}
A.~Castellano, F.~Marchesano, L.~Melotti, and L.~Paoloni, ``{The Moduli Space Curvature and the Weak Gravity Conjecture},'' \href{http://arxiv.org/abs/2410.10966}{{\ttfamily arXiv:2410.10966 [hep-th]}}.

\bibitem{Calderon-Infante:2023ler}
J.~Calder\'on-Infante, A.~Castellano, A.~Herr\'aez, and L.~E. Ib\'a\~nez, ``{Entropy bounds and the species scale distance conjecture},'' \href{http://dx.doi.org/10.1007/JHEP01(2024)039}{{\em JHEP} {\bfseries 01} (2024) 039}, \href{http://arxiv.org/abs/2306.16450}{{\ttfamily arXiv:2306.16450 [hep-th]}}.

\bibitem{borisenko2001isometric}
A.~A. Borisenko, ``Isometric immersions of space forms into {R}iemannian and pseudo-{R}iemannian spaces of constant curvature,'' {\em Russian Mathematical Surveys} {\bfseries 56} no.~3, (2001) 425.

\bibitem{borel2006compactifications}
A.~Borel and L.~Ji, ``Compactifications of symmetric and locally symmetric spaces,'' in {\em Lie Theory: unitary representations and compactifications of symmetric spaces}, pp.~69--137.
\newblock Springer, 2006.

\end{thebibliography}\endgroup

\end{document}